# Anisotropies and Homogeneities of Superconducting Properties in Iron-Platinum-Arsenide Ca$_{10}$(Pt$_3$As$_8$)(Fe$_{1.79}$Pt$_{0.21}$As$_2$)$_5$

Qing-Ping D$_{\text{ING}}$,[1,2] Yuji T$_{\text{SUCHIYA}}$,[1] Yue S$_{\text{UN}}$,[1] Toshihiro T$_{\text{AEN}}$,[1] Yasuyuki N$_{\text{AKAJIMA}}$,[1,2] and Tsuyoshi T$_{\text{AMEGAI}}$[1,2]

[1]*Department of Applied Physics, The University of Tokyo, 7-3-1 Hongo, Bunkyo-ku, Tokyo 113-8656, Japan*

[2]*JST, Transformative Research-Project on Iron Pnictides (TRIP), 7-3-1 Hongo, Bunkyo-ku, Tokyo 113-8656, Japan*

We report a study on the anisotropy of superconducting properties in single crystalline Ca$_{10}$(Pt$_{4-\delta}$As$_8$)(Fe$_{1.79}$Pt$_{0.21}$As$_2$)$_5$ with $T_c$ ~ 13.6 K. Under a field of 5 Oe, the sample reaches fully-diamagnetic state at ~10 K for $H \parallel c$, and ~8 K for $H \parallel ab$, indicating the presence of slight inhomogeneities. The magnetization measurements reveal fish-tail effect in the hysteresis loop for both $H \parallel c$ and $H \parallel ab$. Averaged critical current densities at low magnetic fields along the $c$-axis and $ab$-plane $J_c^{H//c}$ and $J_c^{H//ab}$ at 5 K are estimated to be 0.9×10$^5$ and 0.7×10$^5$ A/cm$^2$, respectively. Resistive transitions under magnetic field show broadening, which is consistent with a relatively large anisotropy of upper critical field. Magneto-optical images reveal homogenous current flow within the $ab$-plane.

KEYWORDS: iron-platinum-arsenide, magneto-optical imaging, critical current density, upper critical field

## 1. Introduction

Stimulated by the discovery of superconductivity in oxypnictide LaFeAsO$_{1-x}$F$_x$ with a critical temperature $T_c \sim 26$ K, a series of iron-based superconductors have been reported in the past several years.[1-18] Recently, iron-platinum-arsenide Ca$_{10}$(Pt$_3$As$_8$)(Fe$_{2-x}$Pt$_x$As$_2$)$_5$ (10-3-8) and Ca$_{10}$(Pt$_4$As$_8$)(Fe$_{2-x}$Pt$_x$As$_2$)$_5$ (10-4-8) have attracted much attention due to its intermetallic PtAs layers as new blocks, which is different from other iron-pnictides.[19-33] While Ca$_{10}$(Pt$_4$As$_8$)(Fe$_{2-x}$Pt$_x$As$_2$)$_5$ (10-4-8) shares a common tetragonal structure with most of iron-pnictides, Ca$_{10}$(Pt$_3$As$_8$)(Fe$_{2-x}$Pt$_x$As$_2$)$_5$ has a triclinic symmetry, which is very rare in superconductors.[20,22] The availability of high-quality single crystals with a controllable amount of Pt doping makes Ca$_{10}$(Pt$_3$As$_8$)(Fe$_{2-x}$Pt$_x$As$_2$)$_5$ particularly attractive.[20-22, 28] Another important feature of Ca$_{10}$(Pt$_3$As$_8$)(Fe$_{2-x}$Pt$_x$As$_2$)$_5$ different from other iron-pnictides is that there is a clear separation of structural instability and superconductivity in the $T$-$x$ phase diagram, which has been supported by transport measurements and direct imaging of structural domains.[21,23] Besides these fundamental properties, current-carrying capability of iron-based superconductor (IBS) is also under intense research. Recently, untextured polycrystalline (Ba$_{0.6}$K$_{0.4}$)Fe$_2$As$_2$ round wires with high transport critical current density ($J_c$) of 0.12 MAcm$^{-2}$ (self field, 4.2 K) and 0.01 MAcm$^{-2}$ (12 T, 4.2 K) has been reported.[34] Such high global transport $J_c$ together with the high upper critical field ($H_{c2}$) make iron-based superconductors very promising for practical applications. Irreversibility field $H_{irr}$, at which the critical current density $J_c$ vanishes, a more accurate parameter to define the application range than $H_{c2}$, is controlled by thermal fluctuations which cause melting of the vortex lattice thus giving rise to dissipation at $H_m(T) \approx 0.005H_{c2}(0)(T_c/T–1)^2/Gi \approx H_{irr}$.[35,36] The strength of thermal fluctuations is quantified by the Ginzburg number $Gi = 2\gamma(k_B T_c m_{ab}/\pi \hbar^2 n \xi)^2$ — the squared ratio of the thermal energy $k_B T_c$ to the superconducting condensation energy in the volume of the Cooper pair, here $n$ is the carrier density. In conventional low-temperature superconductors $Gi \approx 10^{-10}$–$10^{-6}$, and thermal fluctuations are negligible and $H_{irr} \approx H_{c2}$.[35] However, in YBa$_2$Cu$_3$O$_{7-x}$, $Gi \approx 10^{-2}$, $H_{irr}(77\ K) \approx 0.5H_{c2}(77\ K)$, and (Bi,Pb)$_2$Sr$_2$Ca$_2$Cu$_3$O$_x$ with $Gi \approx 1$ has $H_{irr}(77\ K) \ll H_{c2}(77\ K)$. Most IBS have $Gi$ ranging from $\sim 10^{-4}$–$10^{-5}$ to $\sim 10^{-2}$.[37] $Gi$ in Ca$_{10}$(Pt$_3$As$_8$)(Fe$_{2-x}$Pt$_x$As$_2$)$_5$ is estimated to be $\sim 0.1$, which is larger than that in other IBS and YBa$_2$Cu$_3$O$_{7-x}$, and comparable to that in (Bi,Pb)$_2$Sr$_2$Ca$_2$Cu$_3$O$_x$.[38,39] Such high $Gi$ results in very low irreversibility field in Ca$_{10}$(Pt$_3$As$_8$)(Fe$_{2-x}$Pt$_x$As$_2$)$_5$. Together with the very costly platinum, Ca$_{10}$(Pt$_3$As$_8$)(Fe$_{2-x}$Pt$_x$As$_2$)$_5$ may not be suitable for applications. However, the study of critical current density is not limited to application purpose, but it will also elucidate the nature of vortex matter in type-II superconductors.

In this paper, we report on synthesis and characterizations of high-quality Ca$_{10}$(Pt$_3$As$_8$)(Fe$_{2-x}$Pt$_x$As$_2$)$_5$ single crystals. The characterization through X-ray diffraction, EDX, magnetization, resistivity measurements, and magneto-optical (MO) imaging are discussed. Anisotropies in superconducting properties like diamagnetism, upper critical fields, and critical current density and homogeneities have been investigated.

## 2. Experimental Methods

Adopting the method in Ref. 20, Ca chips (Rare Metallic, 99.5%), Pt chips (Furuya Metal, 99.95%), As pieces (Furukawa Denshi, 99.99999%), and FeAs powder were used as starting materials to prepare $Ca_{10}(Pt_3As_8)(Fe_{2-x}Pt_xAs_2)_5$ single crystals. FeAs was prepared by placing stoichiometric amounts of Fe powder (Kojundo Chemical Laboratory, 99%) and As pieces (Furukawa Electric, 99.99999%) in an evacuated quartz tube and reacting them at 1065 $^oC$ for 10 h after heating at 700 $^oC$ for 6 h. The mixture of 2 g with a ratio of Ca: Fe: Pt: As = 23 : 23 : 10 : 44 were loaded into an alumina crucible (inner diameter 12 mm) and sealed in an evacuated quartz tube (inner diameter 18 mm). The whole assembly was heated up to 1100 $^oC$ for 72 hours after heating at 700 $^oC$ for 5 h, followed by cooling down to 1050 $^oC$ at a rate of 1 $^oC$/h. Then the furnace was switched off and cooled to room temperature naturally.

The phase identification of the sample was carried out by means of X-ray diffraction (M18XHF, MAC Science) with Cu-$K\alpha$ radiation generated at 40 kV and 200 mA. The chemical composition of the crystal was confirmed by EDX (S-4300, Hitachi High-Technologies equipped with EMAX x-act, HORIBA). Bulk magnetization is measured by a superconducting quantum interference device (SQUID) magnetometer (MPMS-5XL, Quantum Design). Resistivity measurements were performed in the sample chamber of a SQUID magnetometer by the four-probe method with silver paste for electrical contacts. For MO imaging, a Bi-substituted iron-garnet indicator film is placed in direct contact with the sample, and the whole assembly is attached to the cold finger of a He-flow cryostat (Microstat-HR, Oxford Instruments). MO images were acquired by using a cooled CCD camera with 12-bit resolution (ORCA-ER, Hamamatsu).

## 3. Result and Discussion

The typical dimensions of as-prepared single crystal are $1 \times 1 \times 0.1$ mm$^3$, which are larger than those of $Ca_{10}(Pt_4As_8)(Fe_{2-x}Pt_xAs_2)_5$ crystals. Figure 1(a) shows the X-ray pattern of single crystal. Only 00$l$ reflection peaks are observed, and the interlayer spacing $d$(001) is calculated as 1.0294 nm, which is similar to that in Refs. 20-22. The chemical composition of the crystal determined by EDX analyses is $Ca_{10}(Pt_3As_8)(Fe_{1.79}Pt_{0.21}As_2)_5$. The concentration of Pt in the FeAs layer is calculated from the relative atom ratio of iron and platinum by assuming that there is no Pt vacancy in the $Pt_3As_8$ layer, based on the structure analysis in Refs. 19-22. In an ideal 10-3-8 crystal, the molar ratio of (Ca+As) to (Fe+Pt) is 28/13=2.154, which is distinctly larger than the same ratio of 28/14=2.0 in 10-4-8 phase. The molar ratio of (Ca+As) to (Fe+Pt) in our crystal measured by EDX is 2.169, which is much closer to that in 10-3-8 phase. All these results indicate that our crystals are really 10-3-8 phases. Figure 1(b) shows temperature dependence of the resistivity of $Ca_{10}(Pt_3As_8)(Fe_{1.79}Pt_{0.21}As_2)_5$ single crystal. The resistivity of $Ca_{10}(Pt_3As_8)(Fe_{1.79}Pt_{0.21}As_2)_5$ at room temperature is around 350 $\mu\Omega$ cm, which is almost twice as that in $Ca_{10}(Pt_4As_8)(Fe_{2-x}Pt_xAs_2)_5$, and shows metallic behavior above 80 K and it increases below 80 K, followed by a decrease below 27 K. The resistivity of the parent phase $Ca_{10}(Pt_3As_8)(Fe_2As_2)_5$ is different from those of the parent compound of the other iron pnictides. The resistivity in parent compounds of 1111 and 122 phases behave like poor metal and an anomalous feature in resistivity has been observed at the SDW transition/structural transition.[40,41] However, undoped $Ca_{10}(Pt_3As_8)(Fe_2As_2)_5$ behaves like semiconductor, and the resistivity increases with cooling in the entire temperature range from 300 K to 2 K.[31] This is quite different from that in $Ca_{10}(Pt_4As_8)(Fe_{2-x}Pt_xAs_2)_5$, which shows metallic behavior in the normal state. With Pt substitution into the FeAs layers, metallic behavior appears at high temperatures in $Ca_{10}(Pt_3As_8)(Fe_2As_2)_5$. The upturn at low temperatures becomes less pronounced by Pt substitution. Similar phenomena have also been observed in 1111 phase with element substitution for Fe sites and phosphide superconductor $BaRh_2P_2$.[42-45] Neither Anderson localization theory nor Kondo effect can be used to explain the resistive upturn at low temperatures.[31,43] Further study is needed to clarify this issue. A sharp drop in resistivity is observed starting from 14.2 K, which marks the onset of superconductivity. The resistivity becomes nearly zero at 13.0 K and the transition width is 1.2 K. $T_c$ of our sample is one of the highest, and the normal state resistivity is the lowest among the reports of 10-3-8. Both $T_c$ and the normal state resistivity are very similar to that in Ref. 20. The residual resistivity ratio (RRR) $\rho(300 K)/\rho(T_c^{onset})$ is 1.65. Relatively large $\rho(T_c^{onset})$ and small RRR may be due to the presence of large amount of Pt ions in FeAs layers as suggested in Refs. 19-22. Platinum substitution in FeAs layer is found to be responsible for the lower $T_c$.[32] Band structure calculations indicated that the $Pt_{3+y}As_8$ layers hardly contribute at the Fermi energy, which is supported by angle resolved photoemission experiments, showing that the Fermi-surface topology is similar to other iron pnictides.[21,28] The high critical temperatures in the platinum-iron arsenides are not achieved by Pt substitution inside the iron layers, but by charge doping of FeAs layers. The $T_c$ -x phase diagrams in the 10-3-8 and 10-4-8 compounds are quite different. Pt substitution induces superconductivity at lower temperatures in the 10-3-8 materials, while it is detrimental to $T_c$ in the 10-4-8 compounds, where the FeAs layers are doped by electrons due to a shift of the charge balance between $[(FeAs)_{10}]^{n-}$ and $(Pt_{3+y}As^8)^{m+}$. This interpretation is supported by the observation of superconductivity at 30 K in the electron-doped 10-3-8 compound $(Ca_{0.8}La_{0.2})_{10}(FeAs)_{10}(Pt_3As_8)$ without significant Pt substitution.[32] Temperature dependences of zero-field-cooled (ZFC) and field-cooled (FC) magnetization at $H$=5 Oe for $H \parallel c$ and $H \parallel ab$ of $Ca_{10}(Pt_3As_8)(Fe_{1.79}Pt_{0.21}As_2)_5$ single crystal are shown in Fig.1(c) and (d), respectively. The sample shows an onset of diamagnetism at $T_c$~13.6 K for both directions. The sample reaches full diamagnetism at ~10 K and ~8 K for $H \parallel c$ and and for $H \parallel ab$, respectively. It means that the diamagnetic transition for $H \parallel ab$ is broader than that for $H \parallel c$. This is quite natural since inhomogeneities in each superconducting layer is more faithfully measured in the configuration where magnetic field is applied perpendicular to the short edge of the crystal, $H\perp c$ in the present measurements.

Figure 2 (a) and (b) show the magnetization as a function of magnetic field at several temperatures ranging from 2 to 12.5 K for $H \parallel c$ and $H \parallel ab$, respectively. From these magnetization hysteresis loops, we can evaluate global critical current density $J_c$ for a single crystal using the Bean model with the assumption of field–independent $J_c$. According to the Bean model,[46] $J_c$ [A/cm$^2$] is given by

$$J_c = 20 \frac{\Delta M}{a(1 - a/3b)}. \qquad (1)$$

where $\Delta M$ is $M_{down} - M_{up}$, $M_{up}$ and $M_{down}$ are the magnetization when sweeping field up and down, respectively, $a$ and $b$ are sample widths ($a<b$). Fig. 2(c) and (d) show the magnetic field dependences of $J_c$ calculated from the data shown in Fig. 2(a) and (b) using Eq. (1). A pronounced non-monotonic field dependence of $J_c$ with a broad maximum, fish-tail effect, is clearly observed for $H \parallel c$ as shown in Fig. 2(a). In Fig. 2 (b), the same effect is also observed for $H \parallel ab$, just less pronounced, only appears at temperatures above 7.5 K. It is believed to also exist below 7.5 K with much higher peak field above 50 kOe. The fish-tail effect is more clearly seen in the field dependence of $J_c$ as shown in Fig. 2(c) and 2(d). Crossover from low-field elastic to high-field plastic creep is the most plausible origin of the fish-tail in the $Ca_{10}(Pt_3As_8)(Fe_{1.79}Pt_{0.21}As_2)_5$.[27] This scenario has been partly supported by the strong acceleration of vortex motion above the peak field as demonstrated in various iron-based superconductors.[47,48] In tetragonal two-dimensional systems, there are three kinds of critical currents $J_c^{x,y}$, where $x$ and $y$ indicate the directions of current and magnetic field, respectively. In the case of $H//c$, $J_c^{ab,c}$ is evaluated from the measured magnetization using Eq. (1). For simplicity, we denote $J_c^{ab,c}$ as $J_c^{H//c}$. On the other hand, in the case of $H//ab$, both $J_c^{ab,ab}$ and $J_c^{c,ab}$ contribute to the measured magnetization. Here, we assume that $J_c^{ab,ab}$ is equal to $J_c^{c,ab}$, and obtain weighted average $J_c^{H//ab}$ using Eq. (1). $J_c^{H//c}$ and $J_c^{H//ab}$ at 5 K are estimated to be $0.9\times10^5$ and $0.7\times10^5$ A/cm$^2$, respectively. These $J_c$ values are smaller than that of optimally-doped $Ba(Fe_{1-x}Co_x)_2As_2$ single crystal, but similar to that in $FeTe_{0.6}Se_{0.4}$ and $Ca_{10}(Pt_{4-\delta}As_8)(Fe_{2-x}Pt_xAs_2)_5$ single crystal.[27,49] $J_c^{H//c}$ values in excess of $0.2\times10^5$ A/cm$^2$ are sustained up to 50 kOe even at 5 K, while $J_c^{H//ab}$ values decreased to zero under the field of 35 kOe at 5 K. One of possible reasons for the lower $J_c$ in $Ca_{10}(Pt_{4-\delta}As_8)(Fe_{2-x}Pt_xAs_2)_5$ is that the material may have stacking fault due to Pt deficiency in PtAs layers. In the present case, since $Ca_{10}(Pt_3As_8)(Fe_{1.79}Pt_{0.21}As_2)_5$ is also anisotropic, strong vortex dynamics suppresses $J$ in the measured time window. In both cases, $J_c$ would be strongly suppressed.

The above estimations of $J_c$'s using the Bean model rely on the assumption that homogeneous current is flowing within the sample. To examine this assumption, we made MO imaging of a 46 μm thick $Ca_{10}(Pt_3As_8)(Fe_{1.79}Pt_{0.21}As_2)_5$

single crystal in the remanent state at several temperatures ranging from 5 to 12.5 K. The remanent state is prepared by applying 800 Oe along the *c*-axis for 1 s and removing it after zero-field cooling. Figures 3(a)–(c) show MO images of $Ca_{10}(Pt_3As_8)(Fe_{1.79}Pt_{0.21}As_2)_5$ in the remanent state for $H//c$ at 5, 7.5 and 10 K, respectively. Compared with the MO images of $Ca_{10}(Pt_4As_8)(Fe_{2-x}Pt_xAs_2)_5$, the MO images of the present sample is more clear because larger and more homogenous crystal are easier to get in $Ca_{10}(Pt_3As_8)(Fe_{1.79}Pt_{0.21}As_2)_5$. The maximum flux density is found in the center and along the diagonals. MO images show a nearly uniform current flow in the sample. Figure 3(d) shows profiles of the magnetic induction at different temperatures along the line shown in Fig. 3(a). $J_c$ for a thin superconductor is roughly estimated by $J_c \sim \Delta B/t$, where $\Delta B$ is the trapped field in the full critical state and $t$ is the thickness of the sample. With measured $\Delta B \sim 400$ G at 5 K and $t=46\mu m$, $J_c$ is estimated as $\sim 0.9 \times 10^5$ A/cm$^2$, which is in good agreement with the value obtained from *M-H* curve.

The variations of $T_c$ with magnetic field from 0 to 50 kOe for $H \parallel c$ and $H \parallel ab$ are shown in Figs. 4 (a) and (b), respectively. With increasing field, the resistive transition shifts to lower temperatures accompanied by an increase in the transition width, especially for $H \parallel c$. Such a broadening of resistive transition in magnetic field indicates the effect of strong thermal fluctuations on vortices. The broadening is almost negligible for $H \parallel ab$. With 50 kOe field, the $T_c$ was suppressed to 0.52 $T_{c0}$ for $H \parallel c$ and 0.87 $T_{c0}$ for $H \parallel ab$, where $T_{c0}$ is the transition temperature under zero field.

Figure 4 (c) summarizes the variation of the upper critical field $H_{c2}$ and irreversibility field $H_{irr}$ with temperature for $Ca_{10}(Pt_3As_8)(Fe_{1.79}Pt_{0.21}As_2)_5$. The values of $H_{c2}$ are defined as the field at the midpoint of the resistive transition. The slopes of $H_{c2}$ extracted from the linear part between 10 kOe and 50 kOe are -16.2 kOe/K and -69.0 kOe/K along *c* and *ab* directions, respectively. Using the Werthamer–Helfand–Hohenberg formula $H_{c2}(0)=0.69T_c|dH_{c2}/dT|_{T=T_c}$,[50] which describes the orbital-depairing field for conventional dirty type II superconductors, the value of $H_{c2}$ at $T=0$ K is estimated as $H_{c2}^c(0) = 153$ kOe and $H_{c2}^{ab}(0) = 651$ kOe. The $H_{c2}$ values are smaller than that in $Ca_{10}(Pt_4As_8)(Fe_{2-x}Pt_xAs_2)_5$, which is mainly due to the lower $T_c$ in $Ca_{10}(Pt_3As_8)(Fe_{1.79}Pt_{0.21}As_2)_5$. The anisotropy parameter $\gamma = H_{c2}^{ab}/H_{c2}^c = \sqrt{m_c^*/m_{ab}^*}$ shown in the inset of Fig. 4 is around 10 near $T_{c0}$. However, it should decrease at lower temperatures at least down to the ratio of above-mentioned $H_{c2}$ slopes, which is about 4.3. The anisotropy of $Ca_{10}(Pt_3As_8)(Fe_{1.79}Pt_{0.21}As_2)_5$ is larger than that in most of other iron-based superconductors, but similar to that in another iron-platinum-arsenide $Ca_{10}(Pt_{4-\delta}As_8)(Fe_{2-x}Pt_xAs_2)_5$.[22,27,31] The large anisotropy together with the broadening of the superconducting transition under magnetic field suggests the existence of a wide vortex-liquid phase when the magnetic field is applied along the *c*-axis. There have been several reports showing the presence of reversible region, which corresponds to the vortex liquid phase in a global sense.[51-53] In other words, the separation of $H_{irr}$ and $H_{c2}$ is the evidence for the presence of vortex liquid phase.

In order to extract the superconducting parameters, we have used the Ginzburg–Landau (GL) formula for the coherence length ($\xi$). $\xi$ is calculated from the estimated $H_{c2}(0)$ data using the relations $H_{c2}^{ab}(0) = \Phi_0 / 2\pi\xi_{ab}\xi_c$, $H_{c2}^{c}(0) = \Phi_0 / 2\pi\xi_{ab}^2$, where $\Phi_0 = 2.07 \times 10^{-7}$ G cm$^2$. The obtained values are $\xi_{ab} \sim 1.2$ nm and $\xi_c \sim 4.9$ nm. The values of $H_{irr}$ are defined as the field at 1% of the normal state resistivity. The temperature dependence of the irreversibility field is well approximate by $H_{irr}^{ab}(T) \approx (1-T/T_c)^{1.69}$ and $H_{irr}^{c}(T) \approx (1-T/T_c)^{1.85}$. A thermally-activated flux creep model[54] can be used to explain such a $(1-T/T_c)^n$ type temperature dependence of $H_{irr}$.

Figures 5(a)-(c) reveal the penetration of magnetic flux at 5 K in $Ca_{10}(Pt_3As_8)(Fe_{1.79}Pt_{0.21}As_2)_5$. Starting from the corners of the rectangular sample, so-called discontinuity lines (marked by broken yellow lines in (c)) are formed which cannot be crossed by vortices. These discontinuity lines are sustained with increasing field. Schematics of current distribution and discontinuity lines are presented in Fig. 5 (d). The anisotropy of the in-plane current densities can be easily evaluated from the MO images by measuring the angles of the discontinuity line θ or the flux penetration length $d_1$ and $d_2$.[55] In the present case, in the ab plane $d_1 = d_2$ and the angle θ is 45º, indicating that the critical current density within ab plane is isotropic, consistent with the four-fold symmetry of the superconducting plane.

## 4. Conclusions

In summary, X-ray diffraction, magnetization, resistivity, and magneto-optical measurements have been performed on high quality $Ca_{10}(Pt_3As_8)(Fe_{1.79}Pt_{0.21}As_2)_5$ single crystals. Under a field of 5 Oe, the sample takes around 3 K from $T_c^{onset}$ to reach full diamagnetism for $H \parallel c$, but 5 K for $H \parallel ab$. The magnetization measurements reveal fish-tail hysteresis loops for both $H \parallel c$ and $H \parallel ab$. Magneto-optical imaging revealed high quality of the sample and nearly homogeneous current flow within ab plane. Upper critical fields obtained by resistive transition are 153 kOe and 651 kOe at zero temperature along c and ab directions, respectively. The anisotropy parameter γ near $T_c$ is around 10, which is larger than most of iron-based superconductors.

Acknowledgements

QD would like to thank the support from National Natural Science Foundation of China (NSFC, No. 51202024).

Figure Captions

FIG. 1. (Color online) (a) X-ray diffraction pattern of as-prepared $Ca_{10}(Pt_3As_8)(Fe_{1.79}Pt_{0.21}As_2)_5$ single crystal. (b) Temperature dependence of in-plane resistivity of $Ca_{10}(Pt_3As_8)(Fe_{1.79}Pt_{0.21}As_2)_5$ single crystal. Temperature dependence of ZFC and FC magnetization in $Ca_{10}(Pt_3As_8)(Fe_{1.79}Pt_{0.21}As_2)_5$ measured at $H$ = 5 Oe for (c) $H \parallel c$ and (d) $H \parallel ab$.

FIG. 2. (Color online) Magnetic field dependence of magnetization in $Ca_{10}(Pt_3As_8)(Fe_{1.79}Pt_{0.21}As_2)_5$ at different temperatures ranging from 2 to 12.5 K for (a) $H \parallel c$ and (b) $H \parallel ab$. Magnetic field dependence of averaged critical current densities for (c) $H \parallel c$ and (d) $H \parallel ab$.

FIG. 3. (Color online) Magneto-optical images in the remanent state after applying 800 Oe along the $c$-direction in $Ca_{10}(Pt_3As_8)(Fe_{1.79}Pt_{0.21}As_2)_5$ at (a) 5, (b) 10, and (c) 15 K. (d) The local magnetic induction profiles at different temperatures taken along the dashed lines in (a).

FIG. 4. (Color online) Magnetic field dependence of in-plane resistivity for (a) $H \parallel c$ and (b) $H \parallel ab$. (c) Temperature dependence of upper critical field and irreversibility field for $H \parallel c$ and $H \parallel ab$ defined by the 50% and 1% of normal state resistivity in $Ca_{10}(Pt_3As_8)(Fe_{1.79}Pt_{0.21}As_2)_5$. The inset shows the anisotropy parameter $\gamma(T) = H_{c2}^{ab} / H_{c2}^{c}$.

FIG. 5. (Color online) Magneto-optical images of flux penetration into a nearly rectangular $Ca_{10}(Pt_3As_8)(Fe_{1.79}Pt_{0.21}As_2)_5$ single crystal at $c$-axis fields of (a) 100 Oe, (b) 200 Oe, and (c) 400 Oe after zero-field cooling down to 5 K. The discontinuity lines are marked with yellow broken lines in (c). (d) Schematics of current distribution and discontinuity lines for a fully penetrated state of an isotropic sample.

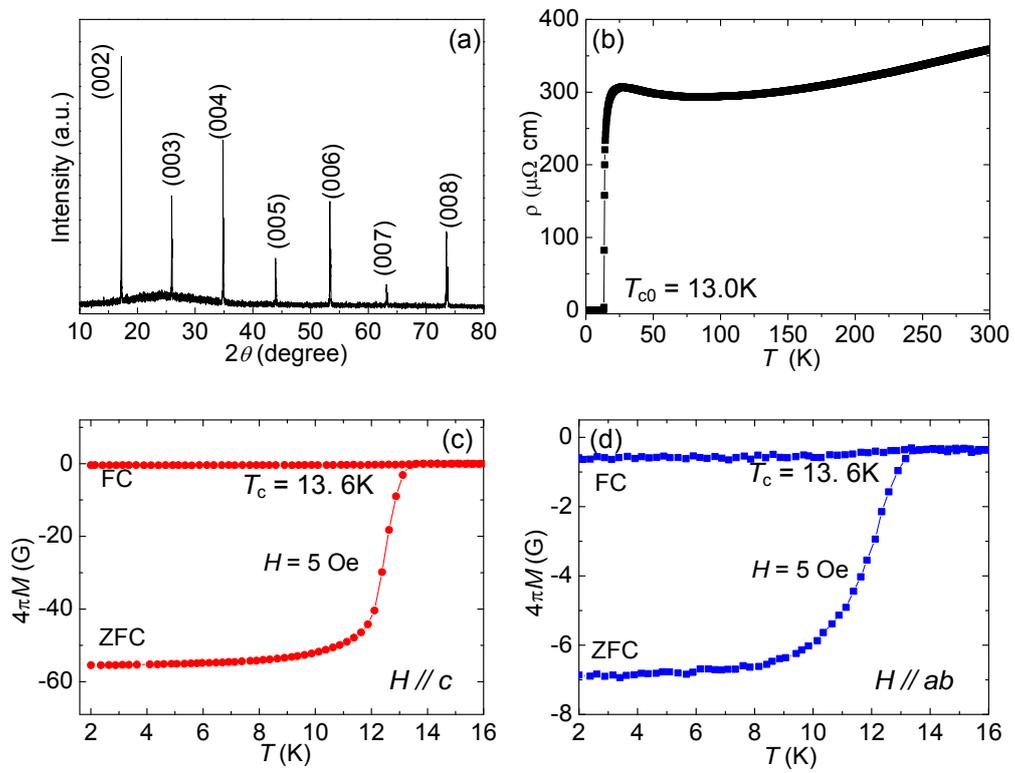

FIG. 1

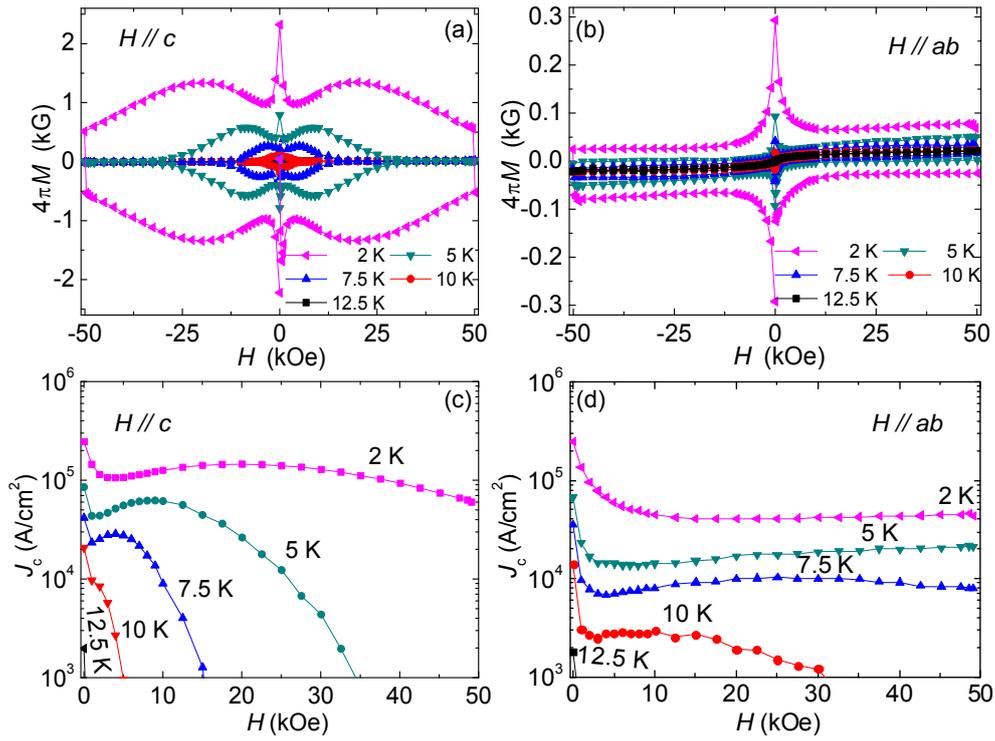

FIG. 2

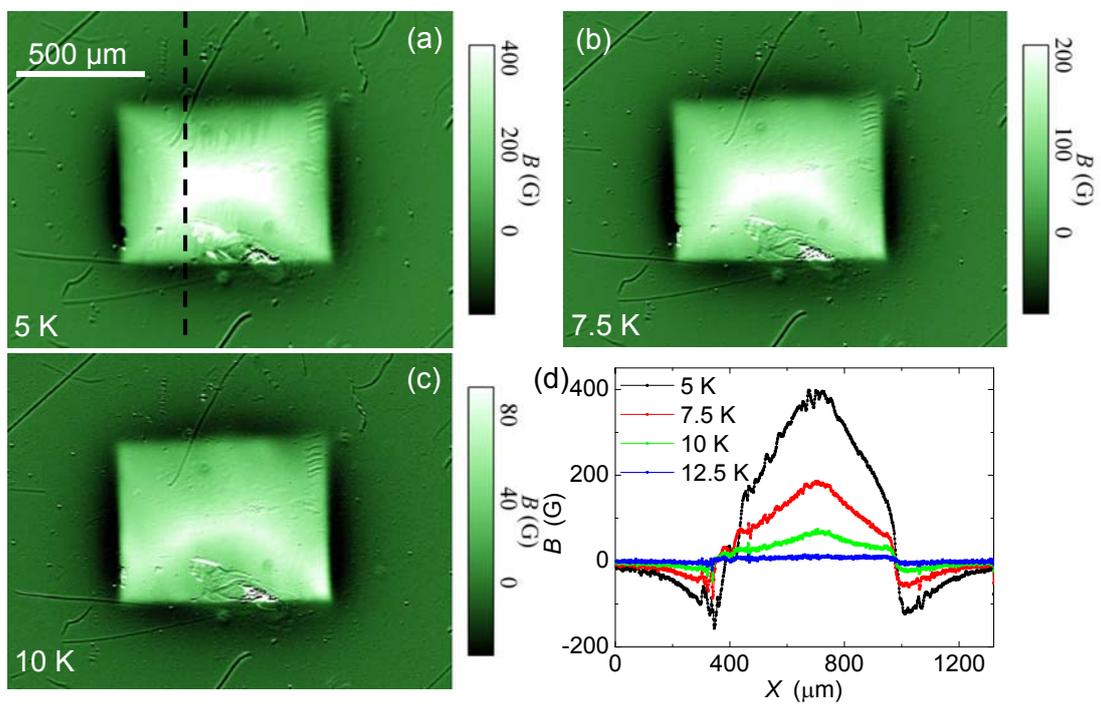

FIG. 3

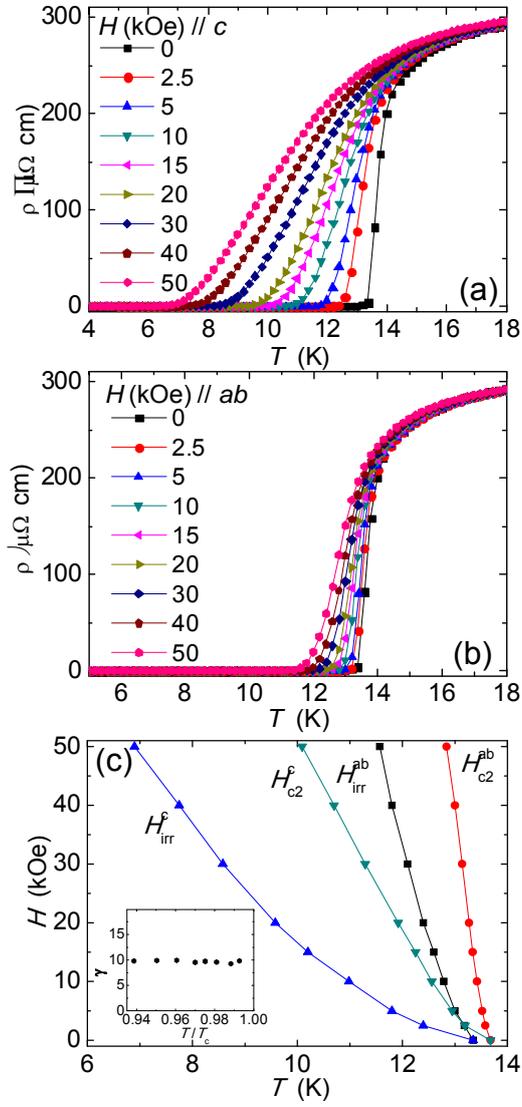

FIG. 4

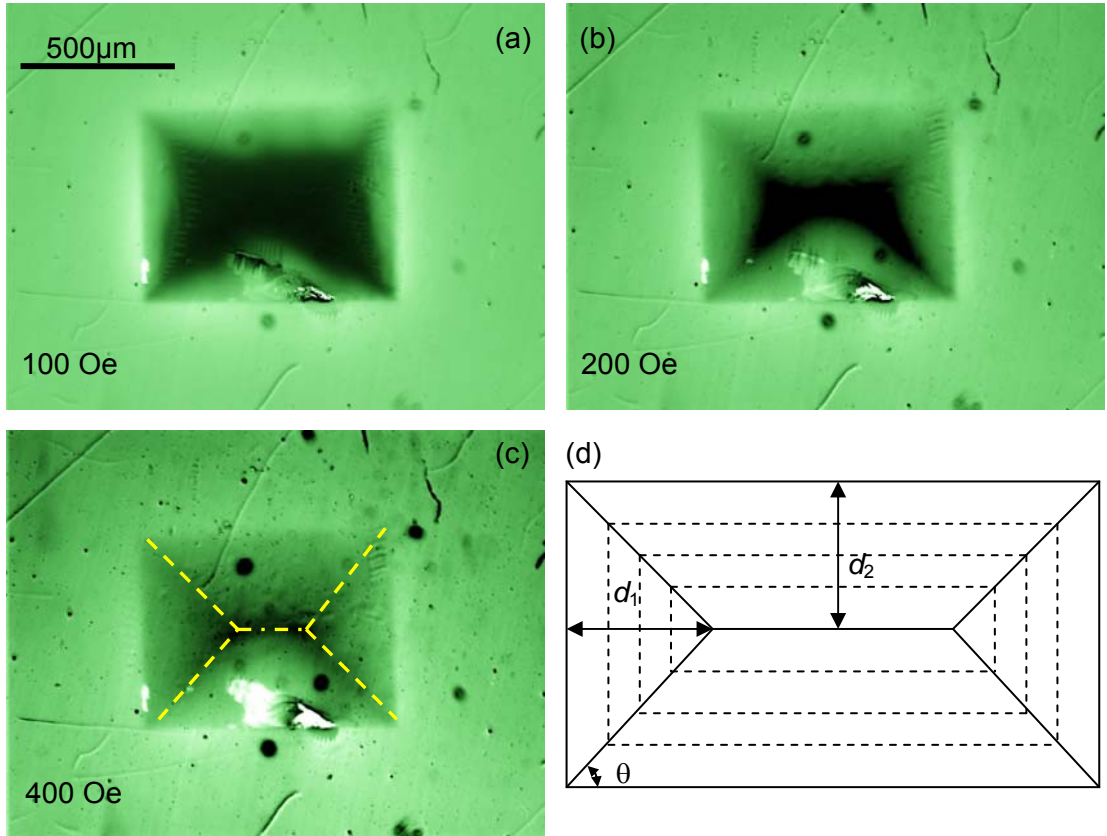

FIG. 5